\documentclass[twocolumn,aps]{revtex4}
\usepackage{graphicx}

\usepackage{graphicx}

\newcommand*{\be}
{\begin{equation}}
\newcommand*{\ee}
{\end{equation}}

\begin{document}
\title{First-order phase transitions in confined systems}
\author{C.A. Linhares}
\address{{\it Instituto de F\'{\i}sica, Universidade do Estado do Rio de Janeiro\\
Rua S\~ao Francisco Xavier, 524, 20559-900 Rio de Janeiro RJ, Brazil}}
\author{A.P.C. Malbouisson, I. Roditi}
\address{{\it CBPF/MCT - 
Rua Dr.~Xavier Sigaud 150, 22290-180, Rio de Janeiro, RJ, Brazil}}

\begin{abstract}
\noindent In a field-theoretical context, we consider the Euclidean 
$(\phi^4+\phi^6)_D$ model  compactified in
one of the spatial dimensions. We are able to determine the
dependence of the transition temperature ($T_{c}$) for a system described by this model, 
confined between two parallel planes, as a function of the distance
($L$) separating them. We show that $T_{c}$ is a concave function of $L^{-1}$. We  determine a minimal separation below which the transition is suppressed.\\
\\
\noindent PACS number(s): 03.70.+k, 11.10.-z
\end{abstract}
\maketitle

In the last few decades, a large amount of work has been done on 
field theoretical models applied to the study of critical phenomena.
In particular, second-order phase transitions have been extensively 
studied in view of the investigation of several material systems.  
An account on the state of the
subject and related topics can be
found, for instance, in Refs. \cite{affleck}-\cite{isaque}.  Questions concerning the existence  of phase transitions may also be raised if one considers the
behaviour of field theories as a function of spatial boundaries.  The
existence of phase transitions would be in this case associated to some
spatial parameters describing the breaking of translational invariance, for
instance, the distance $L$ between planes confining the system.  Studies
of this type have been recently performed \cite{malbouisson2, malbouisson3},
concerning with the spontaneous symmetry breaking in the $\lambda \phi ^4$
theory. In particular, if one considers the Ginzburg--Landau model confined
between two parallel planes, which is assumed to describe a film of some material, the
question of how the critical temperature depends on the thickness $L$ of the
film can be raised. 

Studies on
confined field theory have been done in the literature for a long time. In
particular, an analysis of the renormalization group in finite size
geometries can be found in \cite{zinn, cardy}. These studies have been
performed to take into account boundary effects on scaling laws. In another
related topic of investigation, there are systems that present domain walls
as defects, created for instance in the process of crystal growth by some
prepared circumstances. At the level of effective field theories, in many
cases, this can be modeled by considering a Dirac fermionic field whose mass
changes sign as it crosses the defect, meaning that the domain wall plays
the role of a critical boundary separating two different states of the
system \cite{fosco1, fosco2}.

Under the assumption that information about general features of the behaviour
of systems undergoing phase transitions in absence of external 
influences (like magnetic fields) can be obtained in the
approximation which neglects gauge field contributions in the
Ginzburg--Landau model, investigations have been done with an approach different from the renormalization
group analysis. It has been considered the system confined between two parallel
planes and using the formalism developed in Refs. \cite{malbouisson2,
malbouisson3} it has been investigated how the critical temperature is affected by the
presence of boundaries. In particular a study has been done on how
the critical temperature ($T_c$) of a superconducting film depends on its
thickness $L$\,\cite{luciano,luciano1}.
In this paper we perform a further step, by considering in the same context
 an extended model, which besides the quartic field
self-interaction, a sextic one is also present. It is well known that those
interactions, taken together, lead to renormalizable quantum field theories
in three dimensions and that they are supposed to describe first-order phase transitions.

>From our point of view, as in previous publications, the system to be
studied is a slab of a  material of thickness $L$, the
behaviour of which in the critical region is to be derived from a quantum
field theory calculation of the dependence of the renormalized mass
parameter on $L$. We start from the effective
potential for the theory, which is related to the renormalized mass through
a renormalization condition. This condition, however, reduces considerably
the number of relevant Feynman diagrams contributing to the mass
renormalization, if one wishes to be restricted to first-order terms in both
 coupling constants of the model. In fact, just two diagrams need to be
considered in this approximation: a tadpole graph with the $\phi ^4$
coupling (1 loop) and a ``shoestring" graph with the $\phi ^6$ coupling (2
loops)(see Fig.1). No diagram with both couplings occur. The $L$-dependence appears
from the treatment of the loop integrals, as the  material is
confined between two plane sheets a distance $L$ apart from one another. We therefore take
the space dimension orthogonal to the planes as finite, the other two being
otherwise infinite. This dimension of finite extent is treated in the
momentum space using the formalism of Ref. \cite{malbouisson3}.

We start by stating the Ginzburg-Landau Hamiltonian density in a Euclidean 
$D$-dimensional space, now including both $\phi ^4$ and $\phi ^6$
interactions, in the absence of external fields, given by (in natural units,  
$\hbar =c=k_{B}=1$), 
\begin{equation}
\mathcal{H}=\left| \nabla \varphi \right| ^2+m_0^2\left| \varphi \right|
^2+\frac \lambda 2\left| \varphi \right| ^4+\frac \eta 6\left| \varphi
\right| ^6,  \label{hamiltonian}
\end{equation}
where $\lambda $ and $\eta $ are the (renormalized) quartic and sextic
self-coupling constants, with the bare mass given by $m_0^2=\alpha (T-T_0)$, 
$T_0$ being the bulk transition temperature of the material and 
$\alpha >0$. We consider the system confined between two parallel planes,
normal to the $x$-axis, a distance $L$ apart from one another and use
Cartesian coordinates $\mathbf{r}=(x,\mathbf{z})$, where $\mathbf{z}$ is a ($%
D-1$)-dimensional vector, with corresponding momenta $\mathbf{k}=(k_x,%
\mathbf{q}),\mathbf{q}$ being a ($D-1$)-dimensional vector in momenta space.
The generating functional of Schwinger functions is written in the form 
\begin{equation}
\mathcal{Z}=\int \mathcal{D\varphi }^{*}\mathcal{D\varphi }\exp \left(
-\int_0^Ldx\int d^{D-1}z\,\mathcal{H}\left( \left| \varphi \right| ,\left|
\nabla \varphi \right| \right) \right) ,  \label{partition}
\end{equation}
with the field $\varphi (x,\mathbf{z})$ satisfying the condition of
confinement along the $x$-axis, $\varphi (x\leq 0,\mathbf{z})=\varphi (x\geq
L,\mathbf{z})=$const. Then the field should have a mixed series-integral
Fourier representation of the form 
\begin{equation}
\varphi (x,\mathbf{z})=\sum_{n=-\infty }^\infty c_n\int d^{D-1}q\,b(\mathbf{q%
})e^{-i\omega _nx-i\mathbf{q}\cdot \mathbf{z}}\tilde{\varphi}(\omega _n,%
\mathbf{q}),  \label{fourier}
\end{equation}
where $\omega _n=2\pi n/L$ and the coefficients $c_n$ and $b(\mathbf{q})$
correspond respectively to the Fourier series representation over $x$ and to
the Fourier integral representation over the ($D-1$)-dimensional $\mathbf{z}$%
-space. The above conditions of confinement of the $x$-dependence of the
field to a segment of length $L$ allow us to proceed, with respect to the $x$%
-coordinate, in a manner analogous as is done in the imaginary-time
Matsubara formalism in field theory and, accordingly, the Feynman rules
should be modified following the prescription 
\begin{equation}
\int \frac{dk_x}{2\pi }\rightarrow \frac 1L\sum_{n=-\infty }^\infty ,\qquad
k_x\rightarrow \frac{2n\pi }L\equiv \omega _n.  \label{prescription}
\end{equation}
We emphasize, however, that we are considering an Euclidean field theory in $D$ \emph{%
purely} spatial dimensions, so we are \emph{not} working in the framework of
finite-temperature field theory. Here, the temperature is introduced in the
mass term of the Hamiltonian by means of the usual Ginzburg--Landau prescription.

To continue, we use  some one-loop results described in \cite
{malbouisson2,malbouisson3, ananos}, adapted to our present situation. These results have
been obtained by the concurrent use of dimensional and zeta-function
analytic regularizations, to evaluate formally the integral over the
continuous momenta and the summation over the  frequencies $\omega_{n}$. We get
sums of polar ($L$-independent) terms plus $L$-dependent analytic
corrections. Renormalized quantities are obtained by subtraction of the
divergent (polar) terms appearing in the quantities obtained by application
of the modified Feynman rules and dimensional
regularization formulas. These polar terms are proportional to $\Gamma$-functions having the dimension $D$ in the argument and correspond to the
introduction of counterterms in the original Hamiltonian density. In order
to have a coherent procedure in any dimension, those subtractions should be
performed even for those values of the dimension $D$ for which no poles are
present. In these cases a finite renormalization is performed.

In principle, the effective potential for systems with spontaneous symmetry
breaking is obtained, following the Coleman--Weinberg analysis \cite
{coleman}, as an expansion in the number of loops in Feynman diagrams.
Accordingly, to the free propagator and to the no-loop (tree) diagrams for both
couplings, radiative corrections are added, with increasing number of loops.
Thus, at the 1-loop approximation, we get the infinite series of  1-loop diagrams with
all numbers of insertions of the $\phi ^4$ vertex (two external legs in each
vertex), plus the infinite series of  1-loop diagrams with all numbers of insertions
of the $\phi ^6$ vertex (four external legs in each vertex), plus the
infinite series of 1-loop diagrams with all kinds of mixed numbers of insertions of $
\phi ^4$ and $\phi ^6$ vertices. Analogously, we should include
all those types of insertions in diagrams with 2 loops, etc.
However, instead of undertaking such a daunting computation, even if we
restrict ourselves to the lowest terms in the loop expansion, we remember
that the gap equation we are seeking is given by the renormalization
condition in which the renormalized squared mass is defined as the second
derivative of the effective potential $U(\varphi _0)$ with respect to the
classical field $\varphi _0$, taken at zero field, 
\begin{equation}
\left. \frac{\partial ^2U(\varphi _0)}{\partial \varphi _0{}^2}\right|
_{\varphi _0=0}=m^2.  \label{renorm}
\end{equation}
For our purposes, we do not need to consider
the renormalization conditions for the interaction coupling constants, i.e.,
they may be considered as already renormalized when they are written in the
Hamiltonian above.
At the 1-loop approximation, the contribution of loops with only $\phi ^4$
vertices to the effective potential is  obtained directly from \cite
{malbouisson3}, as an adaptation of the Coleman--Weinberg expression after
compactification in one dimension, 
\begin{eqnarray}
U_1(\phi ,L)&=&\mu ^D\sqrt{a}\sum_{s=1}^\infty \frac{(-1)^{s+1}}{2s}g_1^s\phi
_0^{2s} \nonumber \\
& & \times\sum_{n=-\infty }^\infty \int \frac{d^{D-1}k}{\left( \mathbf{k}%
^2+an^2+c^2\right) ^s}.  
\label{poteffi4}
\end{eqnarray}
In the above formula, in order to deal with dimensionless quantities in the
regularization procedure, we have introduced parameters $c^2=m^2/4\pi ^2\mu
^2$, $a=(L\mu )^{-2}$, $g_1=\lambda /8\pi ^2$ and $\phi _0=\varphi _0/\mu $,
where $\varphi _0$ is the normalized vacuum expectation value of the field
(the classical field) and $\mu$ is a mass scale. The parameter $s$ counts the number of vertices on
the loop.

It is easily seen that only the $s=1$ term contributes to the
renormalization condition (\ref{renorm}). It corresponds to the tadpole
diagram. It is then also clear that all $\phi ^6$-vertex and mixed 
$\phi ^4$- and $\phi ^6$-vertex insertions on the 1-loop diagrams do not contribute
when one computes the second derivative of similar expressions with respect
to the field at zero field: only diagrams with two external legs should 
survive. This is impossible for a $\phi ^6$-vertex insertion at the 1-loop
approximation, therefore the first contribution from the $\phi ^6$ coupling
must come from a higher-order term in the loop expansion.
Two-loop diagrams with two external legs and only $\phi ^4$ vertices are of
second order in its coupling constant, and we  neglect them, as well as
all possible diagrams with vertices of mixed type. However, the 2-loop
shoestring diagram, with only one $\phi ^6$ vertex and two
external legs is a first-order (in $\eta $) contribution to the effective
potential, according to our renormalization criterion.

\begin{figure}[t]
\includegraphics[{height=4.0cm,width=2.5cm,angle=270}]{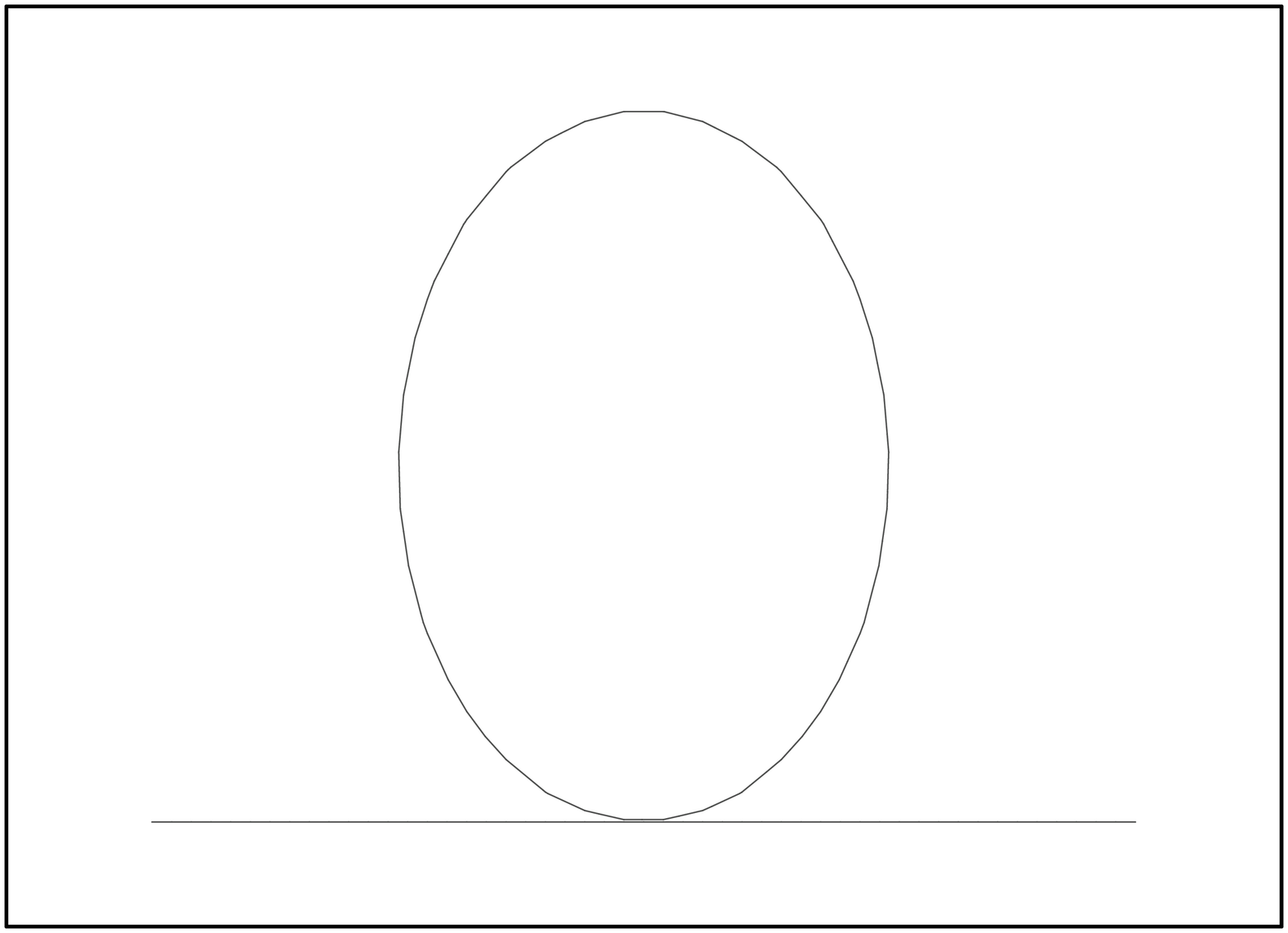}
\includegraphics[{height=4.0cm,width=2.5cm,angle=270}]{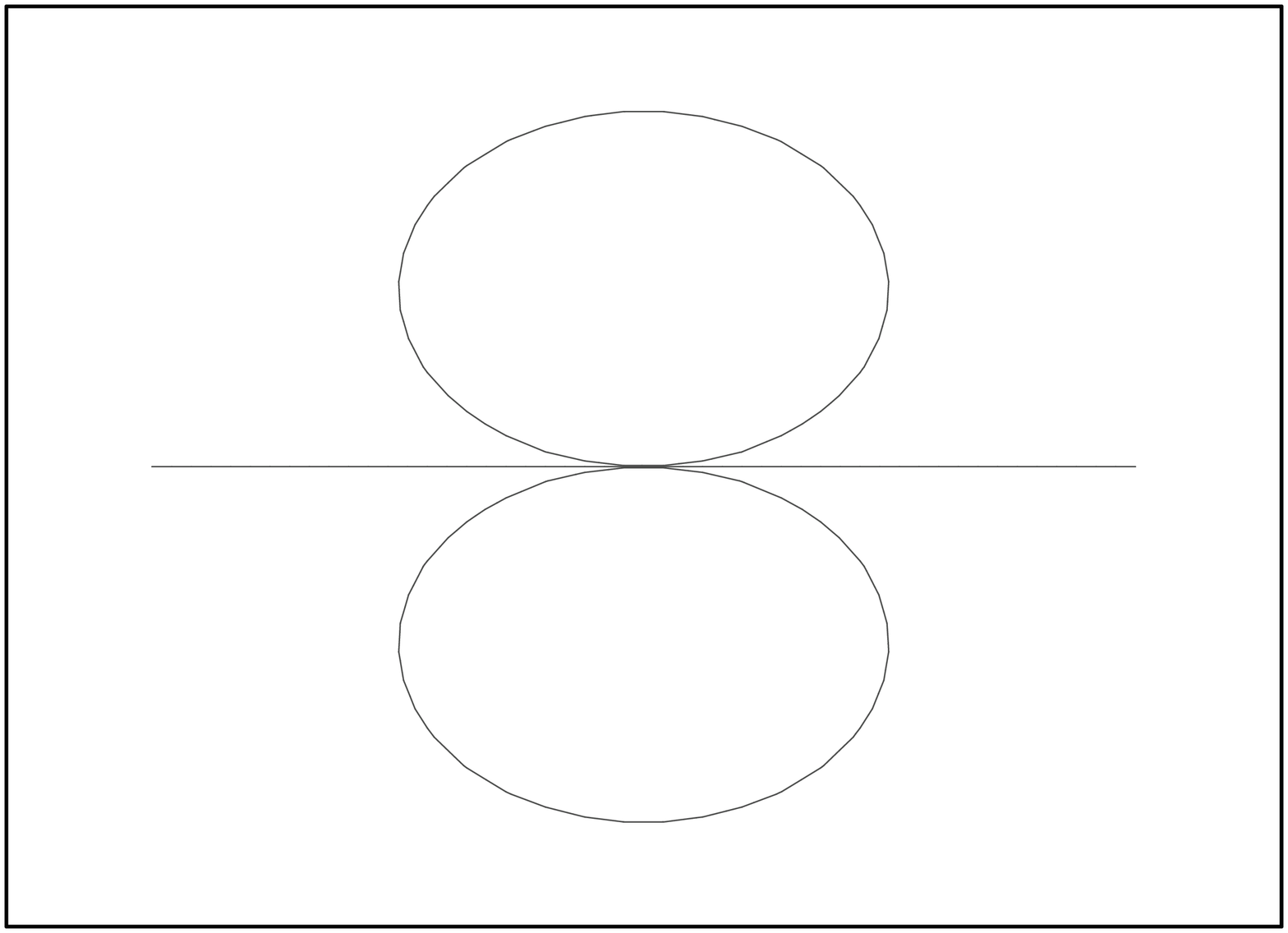}
\caption{Diagrams contributing to the renormalized mass at lowest order in the coupling constants.}
\end{figure}

Therefore the renormalized mass is defined at first-order in both
coupling constants, by the contributions of radiative corrections from only
two diagrams: the tadpole and the shoestring diagrams.
The tadpole contribution reads (putting $s=1$ in eq. (\ref{poteffi4})), 
\begin{equation}
U_1(\phi _0,L)=\mu ^D\sqrt{a}\frac 12g_1\phi _0^2\sum_{n=-\infty }^\infty
\int \frac{d^{D-1}k}{\mathbf{k}^2+an^2+c^2}.  \label{tadpole}
\end{equation}
The integral on the $D-1$ non-compactified momentum variables is 
performed using the dimensional regularization formula 
\begin{equation}
\int \frac{d^dk}{k^2+M}=\frac{\Gamma \left( 1-\frac d2\right) \pi ^{d/2}}{%
M^{1-d/2}};  \label{regdim}
\end{equation}
for $d=D-1$, we obtain 
\begin{eqnarray}
U_1(\phi _0,L)&=&\mu ^D\sqrt{a}\frac{\pi ^{(D-1)/2}}2g_1\phi _0^2\Gamma \left( 
\frac{3-D}2\right) \nonumber \\ 
& &\times\sum_{n=-\infty }^\infty \frac 1{(an^2+c^2)^{(3-D)/2}}. 
\end{eqnarray}
The sum in the above expression may be recognized as one of the
Epstein--Hurwitz zeta-functions, $Z_1^{c^2}(\frac{3-D}2;a)$, which may be
analytically continued to \cite{elizalde} 
\begin{eqnarray}
Z_1^{c^2}(\nu ;a)&=&\frac{2^{\frac{2\nu +1}{2}}\pi ^{\frac{4\nu -1}{2}}}{\sqrt{a}\Gamma (\nu
)}\left[ 2^{\nu -3/2}\left( \frac m\mu \right) ^{1-2\nu }\Gamma \left( \nu
-\frac 12\right) \right.  
\nonumber \\
& & \left. +2\sum_{n=1}^\infty \left( \frac m{\mu ^2Ln}\right) ^{1/2-\nu
}K_{\nu -1/2}(mnL)\right] ,  \nonumber \\
\label{epstein}
\end{eqnarray}
where the $K_\nu $ are Bessel functions of the third kind. The tadpole part
of the effective potential is then
\begin{eqnarray}
U_1(\phi _0,L) &=&\frac{\mu ^Dg_1\phi _0^2}{\left( 2\pi \right) ^{D/2-2}}%
\left[ 2^{-\frac{D+1}{2}}\left( \frac m\mu \right) ^{D-2}\Gamma \left( 1-\frac
D2\right) \right.  \nonumber \\
&&\left. +\sum_{n=1}^\infty \left( \frac m{\mu ^2Ln}\right)
^{D/2-1}K_{D/2-1}(mnL)\right] . \nonumber \\ \label{tadeff}
\end{eqnarray}
Notice  that since we are using dimensional regularization techniques, there is
implicit in the above formulas a factor $\mu ^{4-D}$ in the definition of
the coupling constant $\lambda$. In what follows we make explicit this factor, the
symbol $\lambda $ standing for the dimensionless coupling parameter (which
coincides with the physical coupling constant in $D=4$).

We now turn to the 2-loop shoestring diagram contribution to the effective
potential, using again the Feynman rule prescription for the compactified
dimension. It reads
\begin{equation}
U_2(\phi ,L)=\mu ^{2D-2}a\frac 12g_2\phi _0^2\left[ \frac{\Gamma \left( 
\frac{3-D}2\right) }{\left( 4\pi \right) ^{(D-1)/2}}Z_1^{c^2}\left( \frac{3-D%
}2;a\right) \right] ^2,
\end{equation}
or, after subtraction of the polar term coming from the first term of  Eq.(\ref{epstein}),
\be
U_2^{\rm(Ren)}(\phi ,L)=\frac{\eta\mu^{6-2D}\varphi_0^2}{8(2\pi )^{3D-2}}\left[
 \sum_{n=1}^\infty \left( \frac m{nL}\right) ^{D/2-1}K_{\frac{D-2}{2}}(mnL)\right]^2. \label{eff2ren}
\ee

The renormalized mass with both contributions then satisfies an $L$%
-dependent generalized Dyson--Schwinger equation,
\begin{eqnarray}
m^2(L)&=& m_0^2-\frac{\beta}{\left( 2\pi \right) ^{D/2}}%
\sum_{n=1}^\infty \left( \frac m{nL}\right) ^{D/2-1}K_{\frac{D-2}{2}}(mnL) \nonumber \\
& & +\frac{\rho}{8(2\pi )^{3D-2}}\left[
 \sum_{n=1}^\infty \left( \frac m{nL}\right) ^{D/2-1}K_{\frac{D-2}{2}}(mnL)\right]^2,\nonumber \\
\label{massren1}
\end{eqnarray}
where we have introduced the dimensionful coupling constants $\beta =-\lambda
\mu ^{4-D}$ and $\rho =\eta \mu ^{6-2D}$.

A first-order transition occurs when all the three minima of the potential
\begin{equation}
V(\varphi_{0})=m(L)^2\varphi_0^2-\frac{\beta}{2}\varphi_0^4+\frac{\rho}{6}\varphi_0^6,
\label{potencial}
\end{equation}
where $m(L)$ is the renormalized mass defined above, 
are simultaneously on the line $V(\varphi_{0})=0$. This gives the condition
\begin{equation}
m^2=\frac{3\beta^2}{4 \rho}.
\label{condicao}
\end{equation} 
Notice that the value $m=0$ is excluded in the above condition, for it corresponds to a second-order transition.
For $D=3$, which is the physically interesting situation of the system
confined between two parallel planes embedded in a 3-dimensional Euclidean
space, the Bessel functions entering in the above equations have an explicit form,
$K_{1/2}(z)=\sqrt{\pi }e^{-z}/\sqrt{2z}$, which replaced in Eq.(\ref{massren1}) and 
performing the resulting sum gives
\begin{eqnarray}
m^2(L)&=& \alpha(T-T_0)+\frac{\beta}{\left( 2\pi \right) ^{3/2}}%
\sqrt{\frac{\pi}{2}}\frac{1}{L}\log(1-e^{-mL}) \nonumber \\
& & +\frac{\rho \pi}{8(2\pi )^{7}L^2}
 \left[\log(1-e^{-mL})\right]^2.\nonumber \\
\label{massren2}
\end{eqnarray}
Taking $m=1$ in Eq.(\ref{condicao}), we get the critical temperature,
\begin{eqnarray}
T_c(L)&=& \frac{T_0+1}{\alpha}-\frac{\beta}{\left( 2\pi \right) ^{3/2}}%
\sqrt{\frac{\pi}{2}}\frac{1}{L}\log(1-e^{-mL}) \nonumber \\
& & -\frac{3\beta^2 \pi}{32(2\pi )^{7}L^2}
 \left[\log(1-e^{-mL})\right]^2.
\label{massren3}
\end{eqnarray}

\begin{figure}[t]
\includegraphics[{height=8.0cm,width=5.0cm,angle=270}]{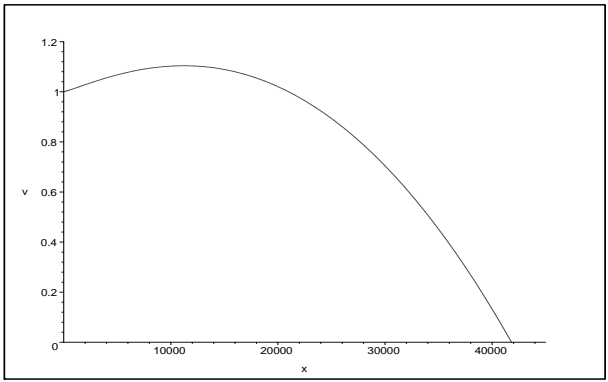}
\caption{Relative critical temperature $v=T_{c}/T_{0}$ as a function of the inverse 
thickness $x=1/L$, in natural units ($m^{-1}$) with $T_{0}=4000,\;\alpha=1,\;\beta=0.1\;\rho=0.3/4$.}
\end{figure}

A plot of the relative critical temperature $T_c/T_0$ against 
the inverse thickness of the system from the above equation, is given in Fig.2 for, in natural units, $\alpha=1{\rm m}^{-1}$,$\beta=0.1{\rm m}^{-1}$ and $T_{0}=4000{\rm m}^{-1}$ ($\approx 10^{\rm o}$K). We see from 
the figure that the critical temperature slightly grows above the bulk transition temperature as the thickness of the system diminishes, reaching a maximum and afterwards starting to decrease until a zero value, corresponding to a minimal allowed thickness for the system.  This behaviour  may be contrasted with the linear decreasing of $T_{c}$ from the maximum value $T_{0}$ with the inverse of the thickness of the system, that has been found for second-order transitions \cite{luciano, urucubaca}. We also remark that in $D=3$, for second-order transitions, one considers $m=0$ and that leads to the need of a pole-subtraction procedure for the mass \cite{isaque}. In our case such a procedure is not necessary, as a first-order  transition must occur for a non-zero value of the mass. This fact, together with the closed formula for the Bessel function for $D=3$, allows us to obtain the exact expression (\ref{massren3}) for the critical temperature.    

This work has received partial financial support from CNPq and Pronex.


\begin{thebibliography}{99}
\bibitem{affleck}  I. Affleck and E. Br\'{e}zin, Nucl. Phys. 257 (1985) 451.

\bibitem{lawrie1}  I.D. Lawrie, Phys. Rev. B 50 (1994) 9456.

\bibitem{lawrie2}  I.D. Lawrie, Phys. Rev. Lett. 79 (1997) 131.

\bibitem{brezin}  E. Br\'{e}zin, D.R. Nelson and A. Thiaville, Phys. Rev. B
31 (1985) 7124.

\bibitem{radzi}  L. Razihovsky, Phys. Rev. Lett. 74 (1995) 4722.

\bibitem{calan}  C. de Calan, A.P.C. Malbouisson and F.S. Nogueira, Phys.
Rev. B 64 (2001) 212502.

\bibitem{malbouisson}  A.P.C. Malbouisson, F.S. Nogueira and N.F. Svaiter,
EuroPhys. Lett. 41 (1998) 547.

\bibitem{halperin} L. Halperin, T. C. Lubensky, S-K. Ma, Phys. Rev. Lett. {\bf 32}, 292(1974).
\bibitem{isaque} L.M. Abreu, A.P.C. Malbouisson, I. Roditi, cond-mat/0305368, to appear in Physica A (2003).

\bibitem{fosco1}  C.D. Fosco and A. Lopez, Nucl. Phys. B 538 (1999) 685.

\bibitem{fosco2}  L. Da Rold, C.D. Fosco and A.P.C. Malbouisson, Nucl. Phys.
B 624 (2002) 485.

\bibitem{zinn}  J. Zinn-Justin, \emph{Quantum Field Theory and Critical
Phenomena} (Clarendon Press, Oxford, 1996), chapter 36.

\bibitem{cardy}  J.L. Cardy (ed.), \emph{Finite Size Scaling} (North
Holland, Amsterdam, 1988).

\bibitem{malbouisson2}  A.P.C. Malbouisson and J.M.C. Malbouisson, J. Phys.
A: Math. Gen. 35 (2002) 2263.

\bibitem{malbouisson3}  A.P.C. Malbouisson, J.M.C. Malbouisson and A.E.
Santana, Nucl. Phys. B 631 (2002) 83.

\bibitem{luciano} L.M. Abreu, A.P.C. Malbouisson, J.M.C. Malbouisson, A.E. Santana, Phys. Rev. B {\bf 67}, 212502 (2003).

\bibitem{urucubaca}  A.P.C. Malbouisson, J.M.C. Malbouisson and A.E. Santana, cond-mat/0205176.

\bibitem{luciano1} A.P.C. Malbouisson, Phys. Rev. B, {\bf 66}, 092502 (2002).


\bibitem{ananos}  G.N.J. A\~{n}a\~{n}os, A.P.C. Malbouisson and N.F.
Svaiter, Nucl. Phys. B 547 (1999) 221.

\bibitem{coleman}  S. Coleman and E. Weinberg, Phys. Rev. D 7 (1973) 1888.

\bibitem{elizalde}  A. Elizalde and E. Romeo, J. Math. Phys. 30 (1989) 1133.


\end{thebibliography}
\end{document}